\documentclass[11pt]{article}

\pdfoutput=1
\usepackage{verbatim}
\usepackage{amssymb}
\usepackage{xcolor}
\usepackage[centertags]{amsmath}
\usepackage[square, comma, sort&compress,numbers]{natbib}	
\usepackage{array,multirow}
\usepackage{graphicx}
\usepackage{bbold}
\usepackage{braket}
\usepackage{wrapfig}
\usepackage{float}
\usepackage{soul}
\usepackage{url}
\usepackage{multicol}
\numberwithin{equation}{section}
\usepackage{tikz}
\usepackage{tikz,pgf}
\usetikzlibrary{shapes}
\usetikzlibrary{calc}
\usetikzlibrary{decorations.pathmorphing}
\usetikzlibrary{decorations.pathreplacing,shapes.misc}
\usetikzlibrary{positioning}

\usetikzlibrary{arrows}
\usetikzlibrary{decorations.markings}
\usetikzlibrary{shadings}
\usetikzlibrary{intersections}
\usepackage{jheppub}
\usepackage{csquotes}
\numberwithin{equation}{section} \makeatletter
\@addtoreset{equation}{section}

\newcommand{\be}{\begin{equation}}
\newcommand{\ee}{\end{equation}}

\makeatletter
\def\@fpheader{\vspace{-.1cm}}
\makeatother

\title{On different approaches to integrable lattice models
}

\author{Vladimir Belavin$^1$, Doron Gepner$^2$, J. Ramos Cabezas$^1$, Boris Runov$^{1,3}$}

\affiliation{$^1$Physics Department, Ariel University, Ariel 40700, Israel.}

\affiliation{$^2$Department of Particle Physics and Astrophysics, Weizmann Institute, Rehovot 76100, Israel.}

\affiliation{$^3$Saint Petersburg State University, 7/9 Universitetskaya nab., St. Petersburg, 199034 Russia.}

\emailAdd{vladimirbe@ariel.ac.il, doron.gepner@weizmann.ac.il, juanjose.ramoscab@msmail.ariel.ac.il, b.runov@spbu.ru}

\abstract{

Interaction-Round the Face (IRF) models are two-dimensional lattice models of statistical mechanics defined by an affine Lie algebra and admissibility conditions depending on a choice of representation of that affine Lie algebra. Integrable IRF models, i.e., the models the Boltzmann weights of which satisfy the quantum Yang-Baxter equation, are of particular interest. In this paper, we investigate trigonometric Boltzmann weights of integrable IRF models. By using an ansatz proposed by one of the authors in some previous works, the Boltzmann weights of the restricted IRF models based on the affine Lie algebras $\mathfrak{su}(2)_k$ and $\mathfrak{su}(3)_k$ are computed for fundamental and adjoint representations for some fixed levels $k$. New solutions for the Boltzmann weights are obtained. We also study the vertex-IRF correspondence in the context of an unrestricted IRF model based on $\mathfrak {su}(3)_k$ (for general $k$) and discuss how it can be used to find Boltzmann weights in terms of the quantum $\hat{R}$ matrix when the adjoint representation defines the admissibility conditions. 
 }

\begin{document}
\maketitle
\flushbottom

\section{Introduction}

Integrable lattice models have various connections with two-dimensional conformal field theories (CFTs). It is known that at criticality, lattice models are described by CFTs~\cite{Belavin:1984vu}. 
The well-studied examples of this relationship include the Ising model, the Yang-Lee edge singularity, the tricritical Ising model, the three-state Potts model, the eight-vertex SOS model, etc., for a review, see \cite{DiFrancesco:1997nk}.

In \cite{Date:1987zz,jimbo1987solvable,Jimbo:1987mu}, Jimbo et al. introduced a class of Interaction-Round the Face (IRF) models that generalize the eight-vertex SOS model \cite{Andrews:1984af}. It was discovered that these IRF models, based on the affine Lie algebras $\mathfrak{su}(n)_k$ in a special regime\footnote{More precisely, the regime III and one-dimensional limit, for details see \cite{Jimbo:1987mu}.}
are associated with minimal models corresponding to the GKO coset $\frac{\mathfrak{su}(n)_k \oplus \mathfrak{su}(n)_1}{\mathfrak{su}(n)_{k+1}}$ \cite{Goddard:1986ee}, in the sense that the characters (branching functions) of this coset coincide with the one-dimensional configuration sums of the IRF models\footnote{See \cite{Gepner:2020ajg, Belavin:2021uzv, Ramos:2022iun} for recent developments regarding this idea.}. 

In~\cite{Gepner:1992kx}, an explicit method for constructing the Boltzmann weights (BWs) of IRF models based on their relationships with CFTs was proposed. The key idea was that the BWs can be expressed in terms of the matrix elements of the braiding matrices of CFTs.

Two other approaches can be used to find BWs of IRF models: the fusion procedure (see, e.g., \cite{Date:1986ju, Jimbo:1987ts}) and the Vertex-IRF correspondence. In the first part of the paper (sections \ref{description},\ref{CFTapproach}), we  
focus on the approach proposed in~\cite{Gepner:1992kx} in the case of restricted IRF models. Then, in section \ref{virfc}, we discuss the Vertex-IRF correspondence in the context of unrestricted IRF models.  
All these approaches
can be combined to analyze more complex IRF models. We discuss this idea in section \ref{conclusions}.

We will define restricted IRF models and explain the key concepts associated with them in section \ref{description}. 
To present our results, below we give a brief overview. Our study focuses on two IRF models based on the affine Lie algebras $\mathfrak{su}(2)_k$ and $\mathfrak{su}(3)_k$. The models are defined on a square lattice, where one of the dominant integral representations of the corresponding algebra sits in each vertex. Two ordered neighboring vertices $a$ and $b$ are considered admissible if the fusion coefficient $N_{a,h}^b$ satisfies the condition 
\begin{equation} \label{condition1}
    N_{a, h}^b>0\;,
\end{equation}
where $N_{a, h}^b$ arises in the tensor product 
\begin{equation} \label{tensorp1}
    a \otimes h = \bigoplus_{i}N_{a,h}^{i} \times [i]\;.
\end{equation}
Here $h$ is some fixed representation, which, together with the algebra itself, defines the IRF model, and $[i]$ denotes the representations appearing in this tensor product.

In section {\ref{CFTapproach}}, we review the ansatz proposed in~\cite{Gepner:1992kx}. In section \ref{su22}, we first consider a warm-up example applying this ansatz to the model based on $\mathfrak{su}(2)_k$ at level $k=2$ with $h$ corresponding to the fundamental representation. This model was studied by Jimbo et al., \cite{ Jimbo:1987mu}. At this level, there are eight possible configurations yielding nonzero BWs. We recover the solutions of  \cite{ Jimbo:1987mu}.

In section \ref{su24}, we study the model based on $\mathfrak{su}(2)_k$ at level $k=4$, with $h$ corresponding to the adjoint representation. This model was investigated by Tartataglia and Pearce in \cite{Tartaglia:2015jla}. At level $k=4$, there are 33 possible configurations of the BWs. Using the ansatz mentioned above \cite{Gepner:1992kx}, we obtain solutions parameterized by four independent parameters that reproduce the solutions found in \cite{Tartaglia:2015jla}.

In section~\ref{su32}, we consider the model based on $\mathfrak{su}(3)_k$ at level $k=2$, with $h$ corresponding to the adjoint representation. At this level, the fusion rules of the algebra are free of multiplicities, and there are 21 possible configurations of the BWs. The solutions we found completely follow the ansatz \cite{Gepner:1992kx}.

In the case of $\mathfrak{su}(3)_k$ algebra  at level $k\geq 3$, the multiplicities in the tensor products are present ($   N_{a, h}^b \geq2 $) and the approach \cite{Gepner:1992kx} is not applicable. In this respect, in section~\ref{virfc}, we explore the vertex-IRF correspondence approach, which is based on the $\hat{R}$ matrix of quantum algebra $\mathfrak{ sl}_q(3)$. 
Using this correspondence, one can express the BWs of unrestricted IRF models in terms of $\hat{R}$ matrix elements.
In the present paper, we describe how the computation of the quantum $\hat{R}$ matrix can be performed along the lines of the method outlined in~\cite{Jimbo:1985ua}, and provide the necessary ingredients for this method, which are the generators of $\mathfrak{sl}_q(3)$ in the adjoint representation. Because the computation of the $\hat{R}$ matrix is bulky due to the large size of the
matrix, the explicit derivation of the BWs will be considered in a separate work \cite{OndiffII}.

In section \ref{conclusions}, we summarize our results and briefly discuss possible applications to a class of  IRF models based on a quotient $\frac{SU(n)_k}{Z_n}$ \cite{Baver:1996kf}. The case $n=2$ has already been studied in \cite{Baver:1996mw}. For $n>2$, the problem of multiplicity arises, and we expect that the present research results will help find exact solutions for this class of IRF models.

\section{Description of the restricted models}  \label{description}

\begin{figure}[H]
    \centering

\begin{tikzpicture}[scale=0.5]
  \foreach \y in {1,2,3,4,5,6}
    \draw (0,\y) -- (7,\y);
  
  \foreach \x in {1,2,3,4,5,6}
    \draw (\x,0) -- (\x,7);

  \foreach \x in {1,2,3,4,5,6}
    \foreach \y in {1,2,3,4,5,6}
      \fill (\x,\y) circle (2pt);

  \draw[red, thick] (3,3) rectangle (4,4);

  \node at (2.7,4.3) {$a$};
  \node at (4.3,4.3) {$b$};
  \node at (4.3,2.7) {$c$};
  \node at (2.7,2.7) {$d$};
  \node at (3.5,3.5) {$u$};
\end{tikzpicture}
    \caption{Two-dimensional lattice.}
    \label{fig1}
\end{figure}
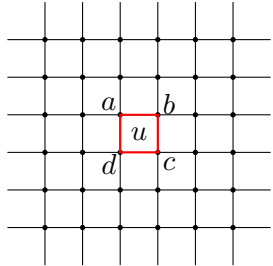

The restricted IRF models are defined on a two-dimensional square lattice as shown in Figure~\ref{fig1}. The fluctuating variables are associated with the lattice vertices. A \textit{face} is formed by four nearest neighboring vertices (in Figure~\ref{fig1} is shown a face with vertices $a, b, c,$ and $d$ ).

IRF models can be defined for a general CFT model, with chiral algebra  $\mathcal{O}$. In the case of affine Lie algebras, the corresponding CFTs are WZW models \cite{Witten:1983ar, DiFrancesco:1997nk}\footnote{In this work, we focus on IRF models based on WZW models with affine Lie algebras $\mathfrak{su}(2)_k$ and $\mathfrak{su}(3)_k$.}. 
The primary fields of the CFT model are labeled by the representations of the algebra $\mathcal{O}$ and the fluctuating variables of the IRF model correspond to \textit{integrable highest-weight representations}, characterized by \textit{dominant integral highest-weights} (defined below). For affine Lie algebras of rank $n-1$, we denote the set of \textit{fundamental weights} as $\Lambda_0, \Lambda_1, \dots, \Lambda_{n-1}$. Consequently, the fluctuating variables residing on the lattice vertices take values from the set of \textit{dominant integral weights} $P_+(n,k)$, 

\begin{equation} \label{inw}
P_+(n,k)=\left\{  a= \sum_{i=0}^{n-1} a_i\Lambda_i= (a_0,a_1,...,a_{n-1}), \quad \sum_{i=0}^{n-1} a_i=k=\text{level}, \quad a_i \in \mathbf{Z}, \quad a_i\ge0 \right\},
\end{equation}
where $k$ is a positive integer referred to as the \textit{level of weights}, serving as a parameter of the model, and $\mathbf{Z}$ represents the set of integers. Among all the weights (\ref{inw}), we choose two weights, $h$ and $v$, to define the admissibility conditions in both the horizontal and vertical directions of the lattice. For a given CFT algebra $\mathcal{O}$ and fields $h$ and $v$, the corresponding IRF model is denoted as IRF ($\mathcal{O}$, $h$, $v$) following the notation in \cite{Gepner:1992kx}. In the present study, we focus on the cases where $h=v$.

Now, we define the \textit{admissibility conditions} for configurations on the vertices. Considering the tensor product of the elements of $P_+(n,k)$ with $h$, see~(\ref{tensorp1}), an ordered pair $(a,b)$ is called \textit{admissible} if it satisfies (\ref{condition1}). Let $(a,b,c,d)$ be the values of the North-West (NW), NE, SE, and SW corners of a face (e.g., the face shown in Figure \ref{fig1}). The face configuration $(a,b,c,d)$ is called \textit{admissible} if the ordered pairs $(a,b)$, $(a,d)$, $(b,c)$, and $(d,c)$ are all admissible, i.e., if

\begin{equation}
N_{a,h}^b >0, \quad N_{a,h}^d >0, \quad N_{b,h}^c >0, \quad N_{d,h}^c>0.
\end{equation}
For each face configuration, we assign a \textit{Boltzmann weight}
\begin{equation}
 \omega \left(\begin{matrix}  a & b \\ d & c 
\end{matrix}\bigg | u \right) .
\end{equation}
Hence, the BWs depend on the configuration $(a,b,c,d)$ and the \textit{spectral parameter} $u$. We set the Boltzmann weight to zero if the face configuration is not admissible. The partition function of the lattice model is given by
\begin{equation}
Z= \sum_{\text{configurations}} \prod_{\text{faces}} \omega \left(\begin{matrix}  a & b \\ d & c 
\end{matrix}\bigg | u \right) .
\end{equation}
We wish to define the BWs in such a way that the model
will be solvable. Namely, that the transfer matrices will commute for different values of the
spectral parameter. This is guaranteed by the Yang–Baxter equation (YBE),
see, e.g., \cite{Gepner:1992kx}
\begin{equation} \label{ibw}
\sum_g \omega \left( \begin{matrix}  a & b \\ f & g 
\end{matrix}\bigg | u +v\right)  \omega \left( \begin{matrix}  f & g \\ e & d 
\end{matrix}\bigg | u \right) \omega \left( \begin{matrix}  b & c \\ g & d 
\end{matrix}\bigg | v \right) = \sum_g \omega \left( \begin{matrix}  a & g \\ f & e
\end{matrix}\bigg | v\right)  \omega \left( \begin{matrix}  a & b \\ g & c 
\end{matrix}\bigg | u \right) \omega \left( \begin{matrix}  g & c \\ e & d 
\end{matrix}\bigg | u+v \right). 
\end{equation}
In the case of trigonometric solutions (considered here), the Boltzmann weights are parameterized by trigonometric functions. On the other hand, for elliptic solutions, the BWs are parameterized by elliptic theta functions, which depend on the elliptic modulus parameter $p$. One can obtain trigonometric solutions from elliptic solutions by taking the limit $p \rightarrow 0$.\footnote{The elliptic solutions for BWs are given by a ratio of two elliptic theta functions. Therefore taking this limit $p \rightarrow 0$ results in a ratio of trigonometric functions.} The possibility of going in the opposite direction is not apparent. Still, the general procedure suggests that to obtain elliptic solutions from trigonometric ones, it is necessary to make the following substitution:
\begin{equation}
    \sin{u} \rightarrow  2 p^{\frac{1}{8}} \sin{u} \prod_{n=1}^{\infty} (1-2 \cos(2u ) p^{n}+p^{2n})(1-p^{n})=\theta_1(p,u).
\end{equation}

In the following section, will explain the approach employed to derive the BWs of the considered models from the Yang-Baxter equation (\ref{ibw}).

\section{CFT approach} \label{CFTapproach}

In this section, we describe the method proposed in \cite{Gepner:1992kx} to obtain the solutions to YBE for the BWs. This method suggests that BWs of the IRF ($\cal O$, $h,h$) can be computed using the following prescription. We label the $N$ representations appearing in the tensor product $h\otimes h$ as follows
\begin{equation} \label{tensorproh}
    h \otimes h= \bigoplus_{j=0}^{N-1}N_{h,h}^j  \times \phi_j.
\end{equation}
According to the proposal, the solutions of the YBE 
are given by
\begin{equation} \label{bw}
\omega \left( \begin{matrix}  a & b \\ d & c 
\end{matrix}\bigg | u \right) \equiv \sum_{j=0}^{N-1} \tilde {P}^{(j)} \left[ a,b,c,d \right] \rho_{j} (u).
\end{equation}
Here, $\tilde {P}^{(j)} \left[ a,b,c,d \right]$  is  a projector defined as
\begin{equation}  \label{trisol}
\tilde {P}^{(j)} \left[ a,b,c,d \right]= \langle a,b,c | P^{(j)} | a,d,c \rangle = \prod_{l \neq  j} \frac{ B_{b,d} \left[   \begin{matrix}  h & h \\ a & c 
\end{matrix}\right]- \delta_{b,d} \lambda_j}{\lambda_l-\lambda_j}\;,
\end{equation}
where $B_{b,d} \left[   \begin{matrix}  h & h \\ a & c 
\end{matrix}\right]$ represents the braiding matrix (for details see, e.g., \cite{Gepner:1992kx, Moore:1988uz}) associated with the CFT $\cal O$ and $\lambda_{j}$ are the eigenvalues of the braiding matrix, these are given by
\begin{equation} \label{theparameterlambda}
\lambda_{j}=\epsilon_j e^{i \pi (\Delta_{\phi_j}-2 \Delta_{h})},
\end{equation}
where $\epsilon_j=\pm1$ depending on whether $\phi_j$ appears symmetrically or anti-symmetrically in (\ref{tensorproh}) and $\Delta_{h},\Delta_{\phi_j}$ are conformal dimensions of the primary fields corresponding to the representations $h,\phi_{j}$ respectively. The conformal dimension $\Delta_a$ of a primary field corresponding to the representation $a$ is given by the formula 
\begin{equation} \label{wzwcd}
\Delta_{a} = \frac{\left( a, a +2\rho \right)}{2(k+g)},
\end{equation}
where $g$ represents the dual Coxeter number of the algebra $\cal O$ under consideration,  and $\rho$ is the Weyl vector $\rho=  \sum_{i=0}^{n-1} \Lambda_i $. The proposal (\ref{bw}) relies significantly on the functions $\rho_j(u)$, which are defined as follows
\begin{equation}
    \rho_j(u) =  \frac{\prod_{r=1}^j  \sin{(  \zeta_{r-1} -u  )} \prod_{r=j+1}^{N-1}  \sin{(  \zeta_{r-1} +u)}} { \prod_{r=1}^{N-1}  \sin{(  \zeta_{r-1}  )} },
\end{equation}
where the parameters $\zeta_i$ referred to as \textit{crossing parameters} in this proposal, are determined as follows
\begin{equation} \label{crossingp}
    \zeta_i= \frac{\pi ( \Delta_{\phi_{i+1}}- \Delta_{\phi_{i}})}{2}, \quad i=0,1,...,N-2.
\end{equation}
The projectors $\tilde{P}^{(j)}[a,b,c,d]$ satisfy the property $\sum_{j=0}^{N-1}P^{(j)}[a,b,c,d] = \mathbf{1}$, where $\mathbf{1}$ denotes the identity operator. Consequently, the BWs (\ref{bw}) satisfy the condition

\begin{equation} \label{inicondition}
    \omega \left( \begin{matrix}  a & b \\ d & c 
\end{matrix}\bigg | 0 \right)=\delta_{b,d}\;.
\end{equation}

Now, let us explain how we compute the BWs in subsections (\ref{su22}, \ref{su24}, \ref{su32}). In general, the $k$ dependence of the braiding matrix appearing in (\ref{trisol}) is not known explicitly, however, it is clear that for concrete values of $k$ and ($a,b,c,d,h$), 
the braiding matrix is a fixed numerical matrix. Therefore, the matrix elements of the $\tilde{P}^{(j)}$ projector also have fixed numerical values, which we need to find. We substitute the ansatz (\ref{bw}) into YBE (\ref{ibw}) and (\ref{inicondition}), and then find solutions for $\tilde{P}^{ (j )} [a,b,c,d]$ for all possible face configurations.
The possible nonzero $\tilde{P}^{(j)}[a,b,c,d]$ values (for a given ($a,b,c,d$)) correspond to the fields $\phi_j$, which belong to the domain:
\begin{equation} \label{intersection}
(a^* \otimes c) \cap (h \otimes h),
\end{equation}
where $a^*$ represents the conjugate representation of $a$, this allows us to reduce the set of unknowns, see \cite{Baver:1996mw}. The results of these computations are presented in the subsequent sub-sections.

It is worth noting 
that as the level $k$ increases (for nontrivial representations $h$ and especially for $\mathfrak{su}(n\geq 3)_k$), the number of admissible face configurations,  
and consequently the number of equations (\ref{ibw}) becomes very large. This makes the task of finding solutions for the BWs more complicated. However, the outlined method shows its effectiveness in various nontrivial examples. For a general level $k$, to determine the BWs, it is required to first find expressions in closed form for the braiding matrix or the projector $\tilde{P}^{(j)}$, as demonstrated for instance in~\cite{Fuchs:1993xu, Baver:1996mw}.

\subsection{Warm-up example: $\mathfrak{su}(2)_2$} \label{su22}
Let us consider the model based on $su(2)_2$ with $h=\Lambda_0+ \Lambda_1$. At this level $k=2$, we find 8 possible BWs. These BWs correspond exactly to the trigonometric solutions found by Jimbo et al., \cite{Date:1987zz,jimbo1987solvable}. Let us describe them by labeling the fields as follows
\begin{equation}
   2\Lambda_0+ 0\Lambda_1:=0, \quad   1\Lambda_0+ 1\Lambda_1:=1,\quad 0\Lambda_0+ 2\Lambda_1:=2.
\end{equation}
The relevant tensor product rules are,
\begin{equation}
\centering
\begin{array}{ll}
 1\otimes 2= 1,    \\
 1 \otimes 1 = 0 \oplus 2 , &   \\
  2 \otimes 2= 0 . &    \\
\end{array}
\end{equation}
Since the tensor product $1 \otimes 1$ contains two representations, $N=2$. Therefore, we have one crossing parameter given by
\begin{equation}
    \zeta_0= \frac{\pi}{4}:= \zeta.
\end{equation}
With these data, we can now search for solutions to the Yang-Baxter equation (\ref{ibw}) and the initial condition (\ref{inicondition}) in the form of (\ref{bw}). We find the following solutions 
\begin{equation}
    \begin{split}
\begin{array}{ll}
\omega( 0 , 1 , 0 , 1,u)= & \frac{s (\zeta +u)}{s (\zeta)} ,\\
\omega( 0 , 1 , 2 , 1,u)= & \frac{s (\zeta -u)}{s (\zeta)},  \\
\omega( 1 , 0 , 1 , 0,u)= & \frac{1}{2}\frac{s (\zeta -u)}{s (\zeta)} + \frac{1}{2} \frac{s (\zeta +u)}{s (\zeta)},\\
\omega( 1 , 0 , 1 , 2 ,u)=& \frac{1}{2}\frac{s (\zeta -u)}{s (\zeta)} - \frac{1}{2} \frac{s (\zeta +u)}{s (\zeta)} ,\\
\omega( 1 , 2 , 1 , 0,u)= &  \frac{1}{2}\frac{s (\zeta -u)}{s (\zeta)} - \frac{1}{2} \frac{s (\zeta +u)}{s (\zeta)},  \\
\omega( 1 , 2 , 1 , 2,u)= &  \frac{1}{2}\frac{s (\zeta -u)}{s (\zeta)} + \frac{1}{2} \frac{s (\zeta +u)}{s (\zeta)} , \\
\omega( 2 ,1 , 0 , 1,u)= & \frac{s (\zeta -u)}{s (\zeta)} , \\
\omega( 2 , 1 , 2 , 1 ,u)=& \frac{s (\zeta +u)}{s (\zeta)}.  \\
\end{array}
\end{split}
\label{weights_sl22}
\end{equation}
Here and in what follows we use the following notations:
\begin{equation}
\omega \left( a,b,c,d,u \right) : = \omega \left( \begin{matrix}  a & b \\ d & c 
\end{matrix}\bigg | u \right),
\end{equation}
\begin{equation}
s(x):= \sin(x).
\end{equation}
The solutions~\eqref{weights_sl22} coincide with the trigonometric limit of BWs obtained in~\cite{Date:1987zz,jimbo1987solvable}. 

\subsection{$\mathfrak{su}(2)_4$ in the adjoint representation} \label{su24}
In this model, we consider $h=2\Lambda_0+2\Lambda_1$, which corresponds to the (2,2)-fused model studied by Tartaglia and Pearce \cite{Tartaglia:2015jla}. At this level $k=4$, the model contains five fields, which are labeled as follows
\begin{equation}
   4\Lambda_0+ 0\Lambda_1:=0, \quad  3\Lambda_0+ 1\Lambda_1:=1,\quad 2\Lambda_0+ 2\Lambda_1:=2, \quad  \Lambda_0 +3 \Lambda_1:=3, \quad  0\Lambda_0+4\Lambda_1:=4.
\end{equation}
The relevant tensor product rules that are used in the admissibility conditions and when fixing the set of unknowns $\tilde{P}^{(j)}$, are the following
\begin{equation}
\begin{array}{ll}
 2\otimes 1= 1\oplus 3, & 1 \otimes 1= 0 \oplus 2,    \\
 2 \otimes2 = 0 \oplus 2 \oplus4, &   1 \otimes 3= 2 \oplus 4, \\
  2 \otimes 3= 1 \oplus 3, &     3 \otimes 3= 0 \oplus 2,\\
  2 \otimes 4= 2, &   4 \otimes 4 = 0.
\end{array}
\end{equation}
Since $N=3$, there are two crossing parameters 
\begin{equation}
\begin{split}
 & \zeta_0= \frac{\pi(\Delta_2-\Delta_0)}{2} = \frac{\pi}{3}:= \zeta ,\\
& \zeta_1=  \frac{\pi(\Delta_4-\Delta_2)}{2} = 2\zeta.
\end{split}
\end{equation}
With this data, we solve the YBE with the ansatz~\eqref{bw}, taking into account~\eqref{inicondition} and find solutions which are parameterized by four parameters $s_1, s_2, s_3, s_4$. In this case, there are 33 admissible configurations on the faces. The solutions for the BWs are
\begin{equation*}
\begin{array}{ll}
 \omega(0 , 2 , 0 , 2,u)= & \frac{s(\zeta +u) s(2 \zeta +u)}{s(\zeta ) s(2 \zeta )} ,\\
\omega( 0 , 2 , 2 , 2,u)= & \frac{s(\zeta -u) s(2 \zeta +u)}{s(\zeta ) s(2 \zeta )} ,\\
 \omega(0 , 2 , 4 , 2,u)= & \frac{s(\zeta -u) s(2 \zeta -u)}{s(\zeta ) s(2 \zeta )}, \\
 \omega(1 , 1 , 1 , 1 ,u)=& \frac{s(\zeta -u) s(2 \zeta +u)}{2 s(\zeta ) s(2 \zeta )}+\frac{s(\zeta +u) s(2 \zeta +u)}{2 s(\zeta ) s(2 \zeta )}, \\
\omega( 1 , 1 , 3 , 1,u)= & \frac{s(\zeta -u) s(2 \zeta -u)}{2 s(\zeta ) s(2 \zeta )}+\frac{s(\zeta -u) s(2 \zeta +u)}{2 s(\zeta ) s(2 \zeta )}, \\
\omega( 1 , 1 , 1 , 3, u)= & \frac{s(\zeta +u) s(2 \zeta +u)}{4 s_1 s(\zeta ) s(2 \zeta )}-\frac{s(\zeta -u) s(2 \zeta +u)}{4 s_1 s(\zeta ) s(2 \zeta )}, \\
 \omega(1 ,1 , 3 , 3,u)= & \frac{s_2 s(\zeta -u) s(2 \zeta +u)}{s(\zeta ) s(2 \zeta )}-\frac{s_2 s(\zeta -u) s(2 \zeta -u)}{s(\zeta ) s(2 \zeta )}, \\
\omega( 1 , 3 , 1 , 1,u)=& \frac{s_1 s(\zeta +u) s(2 \zeta +u)}{s(\zeta ) s(2 \zeta )}-\frac{s_1 s(\zeta -u) s(2 \zeta +u)}{s(\zeta ) s(2 \zeta )}, \\
\omega( 1 , 3 , 3 , 1,u)= & \frac{s(\zeta -u) s(2 \zeta +u)}{4 s_2 s(\zeta ) s(2 \zeta )}-\frac{s(\zeta -u) s(2 \zeta -u)}{4 s_2 s(\zeta ) s(2 \zeta )}, \\
 \omega(1 , 3 , 1 , 3,u)= & \frac{s(\zeta -u) s(2 \zeta +u)}{2 s(\zeta ) s(2 \zeta )}+\frac{s(\zeta +u) s(2 \zeta +u)}{2 s(\zeta ) s(2 \zeta )} ,\\
\omega( 1 , 3 , 3 , 3,u)= & \frac{s(\zeta -u) s(2 \zeta -u)}{2 s(\zeta ) s(2 \zeta )}+\frac{s(\zeta -u) s(2 \zeta +u)}{2 s(\zeta ) s(2 \zeta )}, \\
 \omega(2 , 0 , 2 , 0,u)= & \frac{s(\zeta -u) s(2 \zeta -u)}{4 s(\zeta ) s(2 \zeta )}+\frac{s(\zeta -u) s(2 \zeta +u)}{2 s(\zeta ) s(2 \zeta )}+\frac{s(\zeta +u) s(2 \zeta +u)}{4 s(\zeta ) s(2 \zeta )} ,\\
\omega( 2 , 0 , 2 , 2,u)= & \frac{s(\zeta -u) s(2 \zeta -u)}{16 s_3 s_4 s(\zeta ) s(2 \zeta )}-\frac{s(\zeta +u) s(2 \zeta +u)}{16 s_3 s_4 s(\zeta ) s(2 \zeta )}, \\
\omega( 2 , 0 , 2 , 4,u)= & -\frac{s(\zeta -u) s(2 \zeta -u)}{8 s_4 s(\zeta ) s(2 \zeta )}+\frac{s(\zeta -u) s(2 \zeta +u)}{4 s_4 s(\zeta ) s(2 \zeta )}-\frac{s(\zeta +u) s(2 \zeta +u)}{8 s_4 s(\zeta ) s(2 \zeta )}, \\
\omega( 2 , 2 , 2 , 0,u)= & \frac{2 s_3 s_4 s(\zeta -u) s(2 \zeta -u)}{s(\zeta ) s(2 \zeta )}-\frac{2 s_3 s_4 s(\zeta +u) s(2 \zeta +u)}{s(\zeta ) s(2 \zeta )}, \\
 \omega(2 , 2 , 0 , 2,u)= & \frac{s(\zeta -u) s(2 \zeta +u)}{s(\zeta ) s(2 \zeta )}, \\
 \omega(2 , 2 , 2 , 2,u)= & \frac{s(\zeta -u) s(2 \zeta -u)}{2 s(\zeta ) s(2 \zeta )}+\frac{s(\zeta +u) s(2 \zeta +u)}{2 s(\zeta ) s(2 \zeta )}, \\
\omega( 2 , 2 , 4 , 2,u)= & \frac{s(\zeta -u) s(2 \zeta +u)}{s(\zeta ) s(2 \zeta )} ,\\
\omega( 2 , 2 , 2 , 4,u)= & \frac{s_3 s(\zeta +u) s(2 \zeta +u)}{s(\zeta ) s(2 \zeta )}-\frac{s_3 s(\zeta -u) s(2 \zeta -u)}{s(\zeta ) s(2 \zeta )}, \\
 \omega(2 , 4 ,  2 ,  0,u)= & \frac{s_4 s(\zeta -u) s(2 \zeta +u)}{s(\zeta ) s(2 \zeta )}-\frac{s_4 s(\zeta -u) s(2 \zeta -u)}{2 s(\zeta ) s(2 \zeta )}-\frac{s_4 s(\zeta +u) s(2 \zeta +u)}{2 s(\zeta ) s(2 \zeta )} ,\\
\omega( 2 , 4 , 2 , 2,u)= & \frac{s(\zeta +u) s(2 \zeta +u)}{8 s_3 s(\zeta ) s(2 \zeta )}-\frac{s(\zeta -u) s(2 \zeta -u)}{8 s_3 s(\zeta ) s(2 \zeta )} ,\\
\omega( 2 , 4 , 2 , 4,u)= & \frac{s(\zeta -u) s(2 \zeta -u)}{4 s(\zeta ) s(2 \zeta )}+\frac{s(\zeta -u) s(2 \zeta +u)}{2 s(\zeta ) s(2 \zeta )}+\frac{s(\zeta +u) s(2 \zeta +u)}{4 s(\zeta ) s(2 \zeta )}, \\
\omega( 3 , 1 , 1 , 1,u)= & \frac{s(\zeta -u) s(2 \zeta -u)}{2 s(\zeta ) s(2 \zeta )}+\frac{s(\zeta -u) s(2 \zeta +u)}{2 s(\zeta ) s(2 \zeta )}, \\
 \omega(3 , 1 , 3 , 1,u)= & \frac{s(\zeta -u) s(2 \zeta +u)}{2 s(\zeta ) s(2 \zeta )}+\frac{s(\zeta +u) s(2 \zeta +u)}{2 s(\zeta ) s(2 \zeta )}, \\
\omega( 3 , 1 , 1 , 3 ,u)= &\frac{s_2 s(\zeta -u) s(2 \zeta +u)}{s(\zeta ) s(2 \zeta )}-\frac{s_2 s(\zeta -u) s(2 \zeta -u)}{s(\zeta ) s(2 \zeta )} ,\\
 \omega(3 , 1 , 3 , 3 , u)= &\frac{4 s_1 s_2^2 s(\zeta +u) s(2 \zeta +u)}{s(\zeta ) s(2 \zeta )}-\frac{4 s_1 s_2^2 s(\zeta -u) s(2 \zeta +u)}{s(\zeta ) s(2 \zeta )} ,\\
\omega( 3 , 3 ,  1 ,  1, u)= & \frac{s(\zeta -u) s(2 \zeta +u)}{4 s_2 s(\zeta ) s(2 \zeta )}-\frac{s(\zeta -u) s(2 \zeta -u)}{4 s_2 s(\zeta ) s(2 \zeta )}, \\
\omega( 3 , 3 , 3 , 1,u)= & \frac{s(\zeta +u) s(2 \zeta +u)}{16 s_1 s_2^2 s(\zeta ) s(2 \zeta )}-\frac{s(\zeta -u) s(2 \zeta +u)}{16 s_1 s_2^2 s(\zeta ) s(2 \zeta )}, \\
 \end{array}
\end{equation*}
\newpage
\begin{equation}
\begin{array}{ll}
 \omega(3 , 3 , 1 , 3,u)= & \frac{s(\zeta -u) s(2 \zeta -u)}{2 s(\zeta ) s(2 \zeta )}+\frac{s(\zeta -u) s(2 \zeta +u)}{2 s(\zeta ) s(2 \zeta )} ,\\
 \omega(3 , 3 , 3 , 3,u)= & \frac{s(\zeta -u) s(2 \zeta +u)}{2 s(\zeta ) s(2 \zeta )}+\frac{s(\zeta +u) s(2 \zeta +u)}{2 s(\zeta ) s(2 \zeta )} ,\\
 \omega(4 ,  2 ,  0 ,  2,u)= & \frac{s(\zeta -u) s(2 \zeta -u)}{s(\zeta ) s(2 \zeta )}, \\
 \omega(4 , 2 , 2 , 2,u)= & \frac{s(\zeta -u) s(2 \zeta +u)}{s(\zeta ) s(2 \zeta )}, \\
\omega( 4 , 2 , 4 , 2,u)= & \frac{s(\zeta +u) s(2 \zeta +u)}{s(\zeta ) s(2 \zeta )}. 
\end{array}
\end{equation}
We find that in the particular case 
\begin{equation}
    s_1= -\frac{3}{2},\quad s_2= -\frac{1}{2},\quad s_3= -\frac{1}{4}, \quad s_4=-\frac{1}{2},
\end{equation}
our solutions reproduce exactly the Tartaglia and Pearce solutions.  

\subsection{$\mathfrak{su}(3)_2$ in the adjoint representation} \label{su32}
In this subsection, we present the 21 BWs of the model based on $su(3)_2$ with $h=0\Lambda_0+\Lambda_1+\Lambda_2=(0,1,1)$. In this case, there are six fields, which we enumerate as follows

\begin{equation}
    (2,0,0):=0, \quad   (1,1,0):=1,  \quad    (1,0,1):=2, \quad  (0,2,0):=3, \quad  (0,0,2):=4, \quad (0,1,1):=5.
\end{equation}
The relevant tensor product decompositions for solving this model are 

\begin{equation}
\begin{array}{ll}
5  \otimes 5 = 0\oplus 5, &  
5 \otimes 0=5, \\
5 \otimes 1= 1\oplus 4, & 2  \otimes 4 = 5, \\
5 \otimes   2=  2 \oplus  3, & 1  \otimes 3 = 5 , \\ 
5 \otimes 3= 2, & 1  \otimes 2 = 0\oplus   5, \\
5 \otimes   4 = 1, & 3  \otimes 4 = 0, \\
 &  4  \otimes 3 = 0.
\end{array}
\end{equation}
At this level, the tensor product $5\otimes 5$ contains only two representations, indicating that we have $N=2$ and therefore a single crossing parameter denoted as $\zeta_0$. According to the formula (\ref{crossingp}), this parameter is expected to be $\frac{3\pi}{10}$. However, our calculation shows  that in order to obtain solutions to (\ref{ibw}), the parameter is actually given by
\begin{equation}
    \zeta_0=\frac{\pi}{5}:=\zeta.
\end{equation} 
It is worth noting the important fact that when computing the crossing parameter, one should be aware of the phenomenon called \enquote{pseudo-conformal field theory} \cite{Gepner:1992kx}. Suppose that the theory is a WZW
theory. Then the conformal dimensions are
(\ref{wzwcd}). Actually, we can describe a related pseudo-conformal theory where 
\begin{equation} \label{newparad}
    \bar \Delta_a=r\Delta_a  \quad \text{ modulo $1$},
\end{equation}
where $r$ is any integer strange to $k+g$. This is
a different conformal field theory with the same fusion rules.
Thus, the crossing parameters $\zeta_i$ become
$\zeta_i r$ modulo one. In our particular example,
$\zeta_0=3\pi/10$. But multiplying by $r=4$ we get $\zeta_0=\pi/5$, so these parameters are consistent with one another. Note that one has to change also the braiding matrices,
basically taking $\sin(x)$ to $\sin(r x)$, where $x$ is the appropriate solution, and similarly
for the elliptic solution.

The solutions we derived depend on three parameters: $s_1, s_2, s_3$. They can be expressed as follows

\begin{equation}
\begin{split}
\begin{array}{ll}
\omega( 0 , 5 , 0 , 5,u)= & \frac{s(\zeta +u)}{s(\zeta )} ,\\
\omega( 0 , 5 , 5 , 5,u)= & \frac{s(\zeta -u)}{s(\zeta )} ,\\
\omega( 1 , 1 , 1 , 1,u)= & \frac{\left(\sqrt{5}-1\right) s(\zeta +u)}{2 s(\zeta )}+\frac{\left(3-\sqrt{5}\right) s(\zeta -u)}{2 s(\zeta )} ,\\
\omega( 1 , 1 , 4 , 1,u)= & \frac{s(\zeta -u)}{s(\zeta )}, \\
\omega( 1 , 1 , 1 , 4,u)= & \frac{s_1 s(\zeta +u)}{s(\zeta )}-\frac{s_1 s(\zeta -u)}{s(\zeta )} ,\\
\omega( 1 , 4 , 1 , 1 ,u)=& \frac{\left(\sqrt{5}-2\right) s(\zeta +u)}{s_1 s(\zeta )}+\frac{\left(2-\sqrt{5}\right) s(\zeta -u)}{s_1 s(\zeta )}, \\
\omega( 1 , 4 , 1 , 4,u)= & \frac{\left(\sqrt{5}-1\right) s(\zeta -u)}{2 s(\zeta )}+\frac{\left(3-\sqrt{5}\right) s(\zeta +u)}{2 s(\zeta )} ,\\
\omega( 2 , 2 , 2 , 2,u)= & \frac{\left(\sqrt{5}-1\right) s(\zeta +u)}{2 s(\zeta )}+\frac{\left(3-\sqrt{5}\right) s(\zeta -u)}{2 s(\zeta )} ,\\
\omega( 2 , 2 , 3 , 2,u)= & \frac{s(\zeta -u)}{s(\zeta )} ,\\
\omega( 2 , 2 , 2 , 3,u)= & \frac{s_2 s(\zeta +u)}{s(\zeta )}-\frac{s_2 s(\zeta -u)}{s(\zeta )}, \\
\omega( 2 , 3 , 2 , 2,u)= & \frac{\left(\sqrt{5}-2\right) s(\zeta +u)}{s_2 s(\zeta )}+\frac{\left(2-\sqrt{5}\right) s(\zeta -u)}{s_2 s(\zeta )}, \\
\omega( 2 , 3 , 2 , 3,u)= & \frac{\left(\sqrt{5}-1\right) s(\zeta -u)}{2 s(\zeta )}+\frac{\left(3-\sqrt{5}\right) s(\zeta +u)}{2 s(\zeta )} ,\\
\omega( 3 , 2 , 2 , 2,u)= & \frac{s(\zeta -u)}{s(\zeta )}, \\
\omega( 3 , 2 ,  3 ,  2,u)= & \frac{s(\zeta +u)}{s(\zeta )} ,\\
\omega( 4 , 1 , 1 , 1,u)= & \frac{s(\zeta -u)}{s(\zeta )}, \\
\omega( 4 , 1 , 4 , 1,u)= & \frac{s(\zeta +u)}{s(\zeta )} ,\\
\omega( 5 , 0 , 5 , 0,u)= & \frac{\left(\sqrt{5}-1\right) s(\zeta -u)}{2 s(\zeta )}+\frac{\left(3-\sqrt{5}\right) s(\zeta +u)}{2 s(\zeta )} ,\\
\omega( 5 , 0 , 5 , 5,u)= & \frac{s_3 s(\zeta +u)}{s(\zeta )}-\frac{s_3 s(\zeta -u)}{s(\zeta )} ,\\
\omega( 5 , 5 , 5 , 0,u)= & \frac{\left(\sqrt{5}-2\right) s(\zeta +u)}{s_3 s(\zeta )}+\frac{\left(2-\sqrt{5}\right) s(\zeta -u)}{s_3 s(\zeta )}, \\
\omega( 5 , 5 , 0 , 5,u)= & \frac{s(\zeta -u)}{s(\zeta )} ,\\
\omega( 5 , 5 , 5 , 5,u)= & \frac{\left(\sqrt{5}-1\right) s(\zeta +u)}{2 s(\zeta )}+\frac{\left(3-\sqrt{5}\right) s(\zeta -u)}{2 s(\zeta )} .\\
\end{array}  
\end{split}
\end{equation}
By imposing some specific conditions on the parameters $s_1, s_2, s_3$, these solutions satisfy the symmetry
\begin{equation}
    \omega \left(\begin{matrix}  a & b \\ d & c 
\end{matrix}\bigg | u \right) = \omega \left(\begin{matrix}  a & d \\ b & c 
\end{matrix}\bigg | u \right).
\end{equation}
This is an example of an IRF model based on $\mathfrak{su}(3)_k$ with $h$
in the adjoint representation. For $k >2$ there are multiplicities in tensor products, we will return to this question in section~\ref{conclusions}.

\section{Vertex-IRF correspondence approach for unrestricted models} \label{virfc}

This section focuses on the Vertex-IRF correspondence approach for unrestricted IRF models. Unlike restricted models, the fluctuating variables in unrestricted models are not limited by the level $k$. Specifically, the fluctuating variables located at the vertices of the lattice can assume values from the set $P$ of integral weights of the algebra. Here we are interested in a model based on $\mathfrak{su}(3)_k$, particularly when the adjoint representation defines the admissibility conditions. For the affine Lie algebra $\mathfrak{su}(3)_k$ the set $P$ can be expressed as follows
\begin{equation} \label{intwe}
P=\left\{  a= \sum_{i=0}^{2} a_i\Lambda_i= (a_0,a_1,a_2), \quad a_i \in \mathbf{Z} \right\}.
\end{equation}
We introduce the weights of the adjoint representation of the corresponding finite algebra $\mathfrak{su}(3)$ as follows
\begin{equation} \label{wofadj}   
\begin{array}{cccc}
     e_1= (1,1), & e_2= (-1,2), & e_3= (2,-1), & e_4= (0,0), \\ e_5 = (0,0), & e_6 = (1,-2), & e_7 = (-2,1), &e_8= (-1,-1).
\end{array}
\end{equation}
Their corresponding affine extensions (at level zero) are
\begin{equation}
\begin{array}{cccc}
     \hat{e}_1= (-2, 1,1), & \hat{e}_2= (-1, -1,2), & \hat{e}_3= (-1, 2,-1), & \hat{e}_4= (0,0,0)_1, \\ \hat{e}_5 = (0,0,0)_2, & \hat{e}_6 = (1,1,-2,), & \hat{e}_7 = (1,-2,1), &\hat{e}_8= (2,-1,-1).
\end{array}
\end{equation}
For the unrestricted IRF model, a pair $(a,b)$ is termed admissible if
\begin{equation} \label{admiconun}
    b=a+\hat{e}_i, \quad \text{for some $i=1,2,...,8.$}
\end{equation}
A face configuration $(a,b,c,d)$ is admissible if the pairs $(a,b)$, $(b,c)$, $(a,d)$, $(d,c)$ are all admissible in the sense (\ref{admiconun}).

According to~\cite{date1989one, jimbo1989introduction, Pasquier:1988nc}, the BWs of unrestricted IRF models can be defined in terms of the matrix elements of the so-called quantum $\hat{R}$ matrix. We call this approach \enquote{vertex-IRF correspondence} because the matrix $\hat{R}$ directly determines the BWs of vertex-type models. As mentioned earlier, our study here involves two main objectives: (1) determining the BWs of the unrestricted IRF model based on $\mathfrak{su}(3)_k$ when the adjoint representation defines the admissibility conditions, and (2) extending the approach of Section~\ref{CFTapproach} to a situation where multiplicities occur by obtaining the exact expressions for the functions $\rho_j(u)$ and $\tilde{P}^{(j)}$. Here, we explain the Vertex-IRF correspondence and the method outlined in \cite{Jimbo:1985ua} for computing the $\hat{R}$ matrix elements. Since the computation of the $\hat{R}$ matrix elements in the adjoint representation is a demanding problem in its own right, we plan to address it separately in~\cite{OndiffII}.
By definition, the quantum $\hat{R}$ matrix satisfies the Yang-Baxter equation in the form
\begin{equation} \label{YBEeq2}
    ( I \otimes  \hat{R}(u)  )  (    \hat{R}(u+v) \otimes I       )  (  I \otimes \hat{R}(v)       ) =  (  \hat{R}(v)\otimes I )   (    I \otimes \hat{R}(u+v) )  ( \hat{R}(u) \otimes I )\;.
\end{equation}
Both sides of the equation (\ref{YBEeq2}) act on the tensor product of three vector spaces: $V_1\otimes V_2\otimes V_3$. $I$ is the identity operator and $\hat{R}$ is defined as follows
\begin{equation}
    \hat{R}= P \cdot R\;,
\end{equation}
where $P$ denotes the transposition $P  x \otimes y =y \otimes x$, and $R$, also called quantum $R$ matrix,\footnote{It is also common to write the YBE (\ref{YBEeq2}) in the form $
  R_{12}(u)   R_{13}(u+v)   R_{23}(v)=    R_{23}(v)   R_{13}(u+v)   R_{12}(u)$, where $R_{ij}(u)$ is an operator acting as $R$ on the $i$th and $j$th components and as identity on the other component.} acts on the tensor product of two vector spaces in terms of its matrix elements as follows
\begin{equation} \label{YBEnew3}
    R(u) (e_i\otimes  e_j) =  \sum_{  m,n} R_{i,j}^{m,n} (u)   \quad  (e_m \otimes e_n).
\end{equation}
We can rewrite~\eqref{YBEeq2} in terms of $R$ for a given sextuplet ($i_1, i_2, i_3, f_1, f_2, f_3$), yielding
\begin{equation} \label{YBEeq4}
  \sum_{s_1, s_2, s_3}   R_{s_2, s_3}^{f_1, f_2}(u) R_{i_1, s_1}^{s_2, f_3} (u+v)R_{i_2, i_3}^{s_3, s_1} (v) =  \sum_{s_1, s_2, s_3}R_{s_2, s_3}^{f_2, f_3}(v)R_{s_1, i_3}^{f_1, s_3}(u+v)R_{i_1, i_2}^{s_1, s_2} (u)\;.
\end{equation}
In the context of IRF models, we are interested in the case $V_1=V_2= V_3=V$  and $V$ is the vector space of the adjoint representation of $su(3)$. However, in this case, obtaining the matrix $R$ is a nontrivial task due to its size. Since the adjoint representation contains eight states, the matrix $R$ becomes a $64 \times 64$ matrix. We explain how one can proceed to compute this matrix $R$, but before let us discuss its relations with the BWs of IRF models. The $R$ matrix has the property
\begin{equation}
    R_{i,j}^{k,l} = 0 \quad \text{if   $e_i+e_j\ne e_k+e_l$.}
\end{equation}
Following  the Vertex-IRF correspondence (see, e.g., \cite{date1989one, jimbo1989introduction}), one can compute the BWs of unrestricted IRF models in terms of the matrix elements of $\hat{R}$ as follows
\begin{equation}
   \omega  \left(\begin{matrix}  a & b \\ d & c 
\end{matrix}\bigg | u \right) =    \begin{cases}  \hat{R}_{i,j}^{k,l}(u) \quad \text{if $b-a=\hat{e}_k$, $c-b=\hat{e}_l$, $d-a= \hat{e}_i$, $c-d=\hat{e}_j$},  \\ 0 \quad \text{otherwise}. \end{cases}
\end{equation}
This relation can be graphically represented as
\begin{equation}
\centering
\begin{tikzpicture}[scale=0.5]
  \draw[black, thick] (3,3) rectangle (5,5);
  \foreach \x in {3,5}
    \foreach \y in {3,5}
      \fill (\x,\y) circle (4pt);
  \node at (2.7,5.3) {$a$};
  \node at (5.3,5.3) {$b$};
  \node at (5.3,2.6) {$c$};
  \node at (2.7,2.6) {$d$};
  \node at (4,4) {$u$};
     \node at (7,  4) {$=$};

      \draw[black, thick] (9,3) rectangle (11,5);
       \foreach \x in {9,11}
    \foreach \y in {3,5}
      \fill (\x,\y) circle (4pt);

      \node at (10,5.3) {$e_k$};
  \node at (11.4,4) {$e_l$};
  \node at (10, 2.65) {$e_j$};
  \node at (8.5,4) {$e_i$};
  \node at (10,4) {$u$};
\end{tikzpicture}   
\end{equation}
The lhs square in the equation represents the BWs of the IRF\footnote{In IRF models, when multiplicities arise in the tensor product rules, additional labels are required on the edges to characterize the BWs effectively. We explain this in Section~\ref{conclusions}.} model, while the rhs square corresponds to the BWs of the Vertex model.

To compute the matrix $R$, one can employ the scheme given by Jimbo \cite{Jimbo:1985ua}. This scheme relies on certain fundamental elements. First, we introduce the $sl_q(3)$ quantum algebra and its generators in the adjoint representation. Two simple roots of the algebra are denoted as $\alpha_a$ ($a=1,2$). Three generators, namely $H_a$, $E_a$, and $F_a$, are associated with each simple root. In the adjoint representation, we found these generators can be written in the following manner. The Cartan generators are given by

\begin{equation}
H_1=   \left(
\begin{array}{cccccccc}
 1 & 0 & 0 & 0 & 0 & 0 & 0 & 0 \\
 0 & -1 & 0 & 0 & 0 & 0 & 0 & 0 \\
 0 & 0 & 0 & 0 & 0 & 0 & 0 & 0 \\
 0 & 0 & 0 & 1 & 0 & 0 & 0 & 0 \\
 0 & 0 & 0 & 0 & -1 & 0 & 0 & 0 \\
 0 & 0 & 0 & 0 & 0 & 2 & 0 & 0 \\
 0 & 0 & 0 & 0 & 0 & 0 & 0 & 0 \\
 0 & 0 & 0 & 0 & 0 & 0 & 0 & -2 \\
\end{array}
\right), \quad
 H_2 = \left(
\begin{array}{cccccccc}
 1 & 0 & 0 & 0 & 0 & 0 & 0 & 0 \\
 0 & 2 & 0 & 0 & 0 & 0 & 0 & 0 \\
 0 & 0 & 0 & 0 & 0 & 0 & 0 & 0 \\
 0 & 0 & 0 & -2 & 0 & 0 & 0 & 0 \\
 0 & 0 & 0 & 0 & -1 & 0 & 0 & 0 \\
 0 & 0 & 0 & 0 & 0 & -1 & 0 & 0 \\
 0 & 0 & 0 & 0 & 0 & 0 & 0 & 0 \\
 0 & 0 & 0 & 0 & 0 & 0 & 0 & 1 \\
\end{array}
\right),
\end{equation}

the raising and lowering generators are
\begin{equation}
E_1=\left(
\begin{array}{cccccccc}
 0 & 1 & 0 & 0 & 0 & 0 & 0 & 0 \\
 0 & 0 & 0 & 0 & 0 & 0 & 0 & 0 \\
 0 & 0 & 0 & 0 & 0 & 0 & 0 & 0 \\
 0 & 0 & 0 & 0 & 1 & 0 & 0 & 0 \\
 0 & 0 & 0 & 0 & 0 & 0 & 0 & 0 \\
 0 & 0 & 1 & 0 & 0 & 0 & \frac{q^2+1}{q} & 0 \\
 0 & 0 & 0 & 0 & 0 & 0 & 0 & \frac{q^2+1}{q} \\
 0 & 0 & 0 & 0 & 0 & 0 & 0 & 0 \\
\end{array}
\right), \quad    
 E_2=  \left(
\begin{array}{cccccccc}
 0 & 0 & 0 & 0 & 0 & 1 & 0 & 0 \\
 0 & 0 & \frac{q^2+1}{q} & 0 & 0 & 0 & 1 & 0 \\
 0 & 0 & 0 & \frac{q^2+1}{q} & 0 & 0 & 0 & 0 \\
 0 & 0 & 0 & 0 & 0 & 0 & 0 & 0 \\
 0 & 0 & 0 & 0 & 0 & 0 & 0 & 0 \\
 0 & 0 & 0 & 0 & 0 & 0 & 0 & 0 \\
 0 & 0 & 0 & 0 & 0 & 0 & 0 & 0 \\
 0 & 0 & 0 & 0 & 1 & 0 & 0 & 0 \\
\end{array}
\right),
\end{equation}

\begin{equation}
F_1= \left(\begin{array}{cccccccc}
 0 & 0 & 0 & 0 & 0 & 0 & 0 & 0 \\
 1 & 0 & 0 & 0 & 0 & 0 & 0 & 0 \\
 0 & 0 & 0 & 0 & 0 & 0 & 0 & 0 \\
 0 & 0 & 0 & 0 & 0 & 0 & 0 & 0 \\
 0 & 0 & 0 & 1 & 0 & 0 & 0 & 0 \\
 0 & 0 & 0 & 0 & 0 & 0 & 0 & 0 \\
 0 & 0 & 0 & 0 & 0 & 1 & 0 & 0 \\
 0 & 0 & \frac{q}{q^2+1} & 0 & 0 & 0 & 1 & 0 \\
\end{array} \right), \quad F_2=  \left(\begin{array}{cccccccc}
 0 & 0 & 0 & 0 & 0 & 0 & 0 & 0 \\
 0 & 0 & 0 & 0 & 0 & 0 & 0 & 0 \\
 0 & 1 & 0 & 0 & 0 & 0 & 0 & 0 \\
 0 & 0 & 1 & 0 & 0 & 0 & \frac{q}{q^2+1} & 0 \\
 0 & 0 & 0 & 0 & 0 & 0 & 0 & 1 \\
 1 & 0 & 0 & 0 & 0 & 0 & 0 & 0 \\
 0 & 0 & 0 & 0 & 0 & 0 & 0 & 0 \\
 0 & 0 & 0 & 0 & 0 & 0 & 0 & 0 \\
\end{array}\right).
\end{equation}
They define the $sl_q(3)$ quantum algebra
\begin{equation}
    \begin{split}
       &  k_a= q^{H_a/2} ,\\ &
       k_a E_a=q E_a k_a, \quad k_a E_b= q^{-1/2}E_bk_a,\\&
       k_aF_a= q^{-1}F_a k_a,\quad k_a F_b= q^{1/2} F_bk_a, \\&
       [k_1, k_2]= [E_1,F_2]=[E_2,F_1]=0,\\&
       [E_a,F_a]=   \frac{k_a^2-k_a^{-2}}{q-q^{-1}}, \\&
       E_a^2 E_b- (q+q^{-1})E_a E_bE_b+E_b E_a^2=0,\\&
       F_a^2 F_b-(q+q^{-1})F_aF_bF_a+F_bF_a^2=0,
    \end{split}
\end{equation}
where $a,b=1,2$ and $a\ne b$.
The states of the adjoint representation (\ref{wofadj}) in the Euclidean basis can be written as follows
\begin{equation}
\begin{array}{ccc}
  e_1= ( 1,0,0,0,0,0,0,0  )^T,  &
 e_2= ( 0,1,0,0,0,0,0,0  )^T, &
 e_3= ( 0,0,0,0,0,1,0,0  )^T,\\
 e_4= ( 0,0,1,0,0,0,0,0  )^T,&
 e_5= ( 0,0,0,0,0,0,1,0  )^T,&
 e_6= ( 0,0,0,1,0,0,0,0  )^T,\\
 e_7= ( 0,0,0,0,0,0,0,1  )^T,&
 e_8= ( 0,0,0,0,1,0,0,0  )^T.
\end{array}
\end{equation}
According to Jimbo's scheme, one needs the following two elements 
\begin{equation}
    k_0=  q^{(H_1+H_2)/2}, \quad  E_0= q^{  (H_1-H_2)/3 } \left(   F_2 F_1-q^{-1} F_1 F_2  \right).    
\end{equation}
In~\cite{Jimbo:1985ua} it was shown that the quantum $R$ matrix (\ref{YBEnew3}, \ref{YBEeq4}) satisfies the following system of linear equations (the solution of which is unique up to an overall factor)
\begin{equation} \label{syseq1}
\begin{split}
   & R(u)   \left(  E_a \otimes  k_a^{-1}  +k_a \otimes E_a \right)= \left(    E_a \otimes k_a +k_a^{-1} \otimes E_a \right) R(u), \\&
      R(u)   \left(  F_a \otimes  k_a^{-1}  +k_a \otimes F_a \right)= \left(    F_a \otimes k_a +k_a^{-1} \otimes F_a \right) R(u) ,\\ &
     [ R(u), H_a \otimes I+  I \otimes H_a]=0,\\&
\end{split}
\end{equation} 
and an important fact is that 
\begin{equation} \label{syseq2}
    R(u) \left(   e^u  E_0 \otimes k_0 +k_0^{-1} \otimes E_0 \right) = \left(    e^{u}    E_0     \otimes k_0^{-1}   +k_0 \otimes E_0\right) R(u).
\end{equation}
From (\ref{syseq2}), it is clear that besides the spectral parameter $u$, the $R$ matrix also depends on the parameter $q$  of the quantum algebra. To compute the $R$ matrix, one needs to put the above generators in~\eqref{syseq1},~\eqref{syseq2} and solve these equations for the matrix elements $R_{i,j}^{k,l}$. The results of this computation will be reported in~\cite{OndiffII}. 

\section{Conclusions and discussion} \label{conclusions}
In this work, we have investigated restricted IRF models based on the affine Lie algebras $\mathfrak{su}(2)_k$ and $\mathfrak{su}(3)_k$ for various levels $k$ and representations $h$ that determine the admissibility conditions of the models. Specifically, we have examined the models $\mathfrak{su}(2)_2$ with $h$ corresponding to the fundamental representation, $\mathfrak{su}(2)_4$ with $h$ corresponding to the adjoint representation, and $\mathfrak{su}(3)_2$ with $h$ also in the adjoint representation. For these models, we have employed a previously proposed approach based on the relationship between CFTs and IRF models, described in Section~\ref{CFTapproach}. This approach allows determining the BWs of restricted IRF models in terms of the braiding matrix of the associated CFT and certain trigonometric functions $\rho_j(u)$ that depend on the spectral parameter $u$ and the crossing parameters $\zeta_j$ associated with the conformal dimensions of the fields (representations) appearing in the tensor product $h \otimes h$. By incorporating this ansatz into the Yang-Baxter equation (\ref{ibw}), we have obtained solutions for the BWs of the abovementioned models. Our solutions for $\mathfrak{su}(2)_2$ and $\mathfrak{su}(2)_4$ reproduce known results \cite{jimbo1987solvable, Tartaglia:2015jla}, while our solutions for $\mathfrak{su}(3)_2$ are new. 

The work is focused on particular fixed $k$ levels, however, it is worth noting that the approach can be extended to general levels. To this end, one must first determine the corresponding braiding matrix \cite{Fuchs:1993xu, Baver:1996mw}. By successfully solving these models, we have demonstrated the efficiency of the approach in addressing new IRF models. 
 
In the second part, we explored the approach based on the Vertex-IRF correspondence. The final objective in this direction is to determine the BWs of the unrestricted model based on $\mathfrak{su}(3)_k$ in terms of the quantum $\hat{R}$ matrix. We discussed the method proposed in~\cite{Jimbo:1985ua} for finding $\hat{R}$ and provided all the elements needed to address this problem. 

The motivation for studying the Vertex-IRF correspondence is combining this approach with the CFT approach. Notice that although the Vertex-IRF correspondence provides BWs for unrestricted models, the specific set of BWs for the restricted models can be obtained as a subset of unrestricted BWs. This is because these BWs from this subset satisfy YBE between themselves, as discussed, e.g., in \cite{Jimbo:1987ra, Kuniba:1991yn}. Since the CFT approach is not applicable when multiplicities are present, we plan to investigate how this method generalizes to such cases. Below we describe some details regarding this matter. For $k\geq5$, the tensor product of the adjoint representation (excluding the $\Lambda_0$ component) $h=\Lambda_1+\Lambda_2=(1,1)$ yields:
\begin{equation} \label{tensormul}
h\otimes h= (0,0)\oplus (1,1)_1 \oplus (1,1)_2\oplus (0,3) \oplus(3,0)\oplus(2,2).
\end{equation}
Here, $(1,1)_1$ and $(1,1)_2$ correspond to the adjoint representation that appears twice (this multiplicity corresponds to the multiplicity of the null weights $e_4, e_5$ in (\ref{wofadj})). Notice that to define a face configuration in these cases completely, it becomes necessary to introduce additional labels on the edges. For example, Figure \ref{fig2} illustrates two configurations where the fields at the southeast vertex are the same, but the edges connected to this vertex may have different labels. The multiplicity of $(1,1)$ representation leads to a complication, as it results in two identical values for (\ref{theparameterlambda}), which is not consistent with the ansatz (\ref{trisol}).
\begin{figure}
\centering
\begin{tikzpicture}[scale=0.5]
  \draw[black, thick] (3,3) rectangle (5,5);
  \foreach \x in {3,5}
    \foreach \y in {3,5}
      \fill (\x,\y) circle (4pt);
  \node at (4,5.3) {$\alpha$};
    \node at (2.3,5.7) { \small  $(0,0)$};
  \node at (5.7,5.7) {\small $(1,1)$};
    \node at (5.5,4) {$\beta_1$};
  \node at (5.7,2.3) {\small $(1,1)_1$};
    \node at (2.4,2.3) {\small $(1,1)$};
    \node at (4 ,2.5) {$\gamma_1$};
  \node at (2.6,4) {$\lambda$};
  \node at (4,4) {$u$};

      \draw[black, thick] (11,3) rectangle (13,5);
       \foreach \x in {11,13}
    \foreach \y in {3,5}
      \fill (\x,\y) circle (4pt);

\node at (10.3,5.7) { \small  $(0,0)$};
  \node at (13.7,5.7) {\small $(1,1)$};

  \node at (13.7,2.3) {\small $(1,1)_2$};
    \node at (10.4,2.3) {\small $(1,1)$};

      \node at (12,5.3) {$\alpha$};
  \node at (13.5,4) {$\beta_2$};
  \node at (12, 2.5) {$\gamma_2$};
  \node at (10.6,4) {$\lambda$};
  \node at (12,4) {$u$};
\end{tikzpicture}  
\caption{}\label{fig2}
\end{figure}

Studying the generalization of the CFT approach is a crucial step in further investigations of IRF models which cannot be solved solely by applying the fusion procedure \cite{Jimbo:1987ts}. Interesting examples are the models based on the algebra of the quotient group $\frac{SU(n)_k}{Z_n}$. These quotients were investigated in the context of non-diagonal modular invariant CFTs \cite{Gepner:1986wi}. Their tensor product rules were computed in \cite{Baver:1996kf}. These theories exhibit novel features not present in the original $\mathfrak{su}(n)_k$ models. For instance, only representations that are $Z_n$-invariant are present, and there are so-called \enquote{fixed-point representations}, which play a significant role. 

To illustrate these features, let us consider an example from \cite{Baver:1996kf}, the quotient $\frac{SU(3)_6}{Z_3}$. The tensor product of two adjoint representations, in this case, is given by
\begin{equation}
(1,1)\otimes (1,1) = (0,0) \oplus2 (1,1) \oplus 2 (3,3) \oplus(2,2)\oplus (2,2)'\oplus (2,2)'',
\end{equation}
on rhs, the first three representations are also present in (\ref{tensormul}), but in this theory, the representations are identified up to the external automorphism $\sigma$, defined as $\sigma(a_1, a_2) = (k-(a_1+a_2), a_1)$. Consequently, the representations $(3,3)$, $(0,3)$, and $(3,0)$ are identified. The fixed-point representations are those invariant under the automorphism $\sigma$, and for this level, they are $(2,2)$, $(2,2)'$, and $(2,2)''$. These fixed-point representations are treated as three distinct fields (representations). In the original $\mathfrak{su}(3)_k$ models, these fixed-point representations are absent, and the question is how to incorporate them in the fusion procedure. Thus, finding the BWs of IRF models based on these quotients remains an open question. In \cite{Baver:1996mw}, the case of $\frac{SU(2)_k}{Z_2}$ was solved by combining the CFT approach and known solutions of the $\mathfrak{su}(2)_k$ model. The strategy involved the approach \ref{CFTapproach} to find the BWs containing fixed-point representations, and for other BWs, the solutions of $\mathfrak{su}(2)_k$ were used. However, if we want to apply this strategy to other cases (such as $\frac{SU(3)_k}{Z_3}$), we encounter the following difficulty. The tensor product rule of fixed-point representations in these cases (see section 2.1 of \cite{Baver:1996kf}) involves multiplicity, unlike the case of $\frac{SU(2)_k}{Z_2}$. Consequently, this prevents us from employing the approach \ref{CFTapproach}. We leave this problem as an open question for future research.

\vspace{1cm}
\textbf{Acknowledgment:} J.R. expresses gratitude to the organizers of the Workshop on Integrability 2023 held at the University of Amsterdam, where this work was presented. The work of B.R. was supported by the Ministry of Science and Higher Education of the Russian Federation (agreement no. 075–15–2022–287).
 
\bibliographystyle{JHEP} 
\bibliography{refs} 
\end{document}